\documentclass[11pt,amstex,aps,amsfonts,prsty,preprint]{article}
\usepackage{graphicx}
\usepackage{dcolumn}
\usepackage{bm}
\usepackage{amssymb}
\usepackage{amsmath}
\usepackage{a4wide}
\usepackage{citesort}

\begin{document}

\begin{titlepage}

\vspace{0.3in}

\begin{center}

{\LARGE \bf Equilibrium Properties of A Monomer-Monomer Catalytic Reaction
on A One-Dimensional Chain}

\vspace{0.5in}

{\Large \bf G. Oshanin$^1$, M. N. Popescu$^{2,3}$, and S. Dietrich$^{2,3}$}

\vspace{0.3in}

{\large \sl $^1$ Laboratoire de Physique Th{\'e}orique des Liquides, \\
Universit{\'e} Paris 6, 4 Place Jussieu, 75252 Paris, France
}

\vspace{0.1in}

{\large \sl $^2$
Max-Planck-Institut f\"ur Metallforschung, \\
Heisenbergstr. 3, D-70569 Stuttgart, Germany
}

\vspace{0.1in}

{\large \sl $^3$ Institut f\"ur Theoretische und Angewandte Physik, \\
Universit\"at Stuttgart, Pfaffenwaldring 57, D-70569 Stuttgart, Germany
}

\vspace{0.5in}

\begin{abstract}

We study the equilibrium properties of a lattice-gas model of an
$A + B \to 0$ catalytic reaction on a one-dimensional chain in contact
with a reservoir for the particles.
The particles of species $A$ and $B$ are in thermal contact with their
vapor phases acting as reservoirs, i.e., they may adsorb onto empty lattice
sites and may desorb from the lattice. If adsorbed $A$ and $B$ particles
appear at neighboring lattice sites they instantaneously react and both desorb.
For this model of a catalytic reaction in the adsorption-controlled limit,
we derive analytically the expression of the pressure and present exact results
for the mean densities of particles and for the compressibilities of the
adsorbate as function of the chemical potentials of the two species.
\end{abstract}

\vspace{0.7in}

\end{center}

\noindent \hspace{.4in}{\bf PACS}: 05.50.+q; 64.60.Cn; 68.43.De; 82.65.+r

\end{titlepage}

\section{Introduction}
During the two decades following the work of Ziff, Gulari, and Barshad (ZGB)
\cite{zgb} there has been a remarkable development in the theoretical analysis
of catalytically activated reactions. The ZGB model, sometimes referred
to as the ``monomer-dimer" model, has been introduced to describe the important
process of oxidation of carbon monoxide on a catalytic surface \cite{1a}.
Within this model, a monomer ($\rm CO$) adsorbs onto a single vacant
site of the surface (since dissociation of $\rm CO$ is not considered,
$\rm CO$ is treated as a monomer), while a dimer ($\rm O_2$) adsorbs onto a
pair of adjacent $vacant$ sites and then immediately dissociates. Both $\rm CO$
and $\rm O_2$ are in thermal contact with their gaseous reservoirs.
Nearest neighbors of adsorbates composed of a dissociated $\rm O$ atom
and a $\rm CO$ molecule react and form a $\rm CO_2$ molecule, which then desorbs
from the metal surface. The ZGB model predicts remarkable new features compared
to the classical mean-field kinetic schemes of rate equations \cite{1a}: for a
two-dimensional catalytic substrate as the $\rm CO$ gas pressure is lowered
the system undergoes a first-order transition from a $\rm CO$  saturated
inactive phase (zero rate of $\rm CO_2$ production) into a reactive steady
state (non-zero rate of $\rm CO_2$ production) followed by a continuous
transition into an $\rm O_2$-saturated inactive phase. This continuous
transition was shown to belong to the same universality class as the directed
percolation and the Reggeon field theory \cite{universality_class}.
A simpler ``monomer-monomer" model, in which particles of both species can adsorb
on single but different sites, has also been proposed \cite{fich,mea,sad,alb}, and
it has been shown that in (2+1) spatial dimensions this system exhibits a first-order
transition from a phase saturated with one species to one saturated with the
other. Allowing desorption of one species leads to a continuous transition
which also belongs to the directed percolation universality class \cite{alb,red}.

These observations have significantly increased the interest in the properties
of such models of catalytically activated reactions. Different aspects of
the dynamics of the adsorbed phase for these two models, as well as for their
extensions to molecules with more complicated structures (e.g., dimer-dimer
\cite{alb2}, dimer-trimer \cite{ben} models), have been examined thoroughly
\cite{con,mai,cle,eva,krap,park,dick,frach,sholl,mon,arg,nic,dic}, and the studies
have confirmed an essentially collective, many-particle behavior.
On the other hand, the equilibrium properties of catalytically activated
reactions in systems in which the reactive species undergo continuous exchanges
with their vapor phases acting as reservoirs, i.e., adsorb on and desorb from
the catalytic substrate, have been much less studied, and, consequently, the
understanding of the equilibrium state remains rather limited.
Only recently such equilibrium properties, revealing a rather non-trivial
behavior, have been obtained for the more simple case of single-species reactions
$A + A \to 0$ on a one-dimensional chain with a random distribution of
catalytic segments or catalytic sites \cite{osh_JPA02,osh_JSP02}. Kinetics of
the diffusion-limited $A + A \to 0$ reactions, controlled by the constraint that
the particles may undergo reactions only when they meet each other
in the vicinity of special catalytic sites has been discussed in Ref.~\cite{blum}.

Here we study the properties of the equilibrium state for a very simple model
of a monomer-monomer $A + B \to 0$ catalytic reaction on a one-dimensional chain.
Within this model the particles of species $A$ and $B$ undergo continuous exchanges
with the mixed vapor phase, i.e., they adsorb onto empty lattice sites on
the chain and may thermally desorb from them, while the vapor phase as a reservoir
is steadily maintained at constant chemical potentials. In addition, if any
adsorbed $A$ and $B$ particles appear at neighboring lattice sites, they
instantaneously react and both leave the chain (i.e., desorb), and the reaction
product $AB$ is completely removed from the system. For this model we derive the
expression  for the pressure of the adsorbed particles, which provides the complete
thermodynamical description of the system, and we present exact asymptotic
(in respect to the chemical potential) results for the mean densities of the particles
and for the compressibilities of the system. We note that one could also consider
a model in which the product $AB$ is not removed from the system but is forming its
own vapor phase from which it may readsorb on the substrate. However, this more
complicated case will not be studied here.

In Sec. 2 we describe the model and introduce basic notations. In Sect. 3 we
derive the recursion relations obeyed by the partition function of the system
and obtain the solution of these recursion relations. The connection of this model
with classical spin $S=1$ models is discussed in Sect. 4. In Sect. 5 we derive the
expressions of the mean densities of the particles and of the compressibilities of
the system, analyze their asymptotic behavior as function of the chemical potentials,
and present results of Monte Carlo simulations for the occupation of the lattice in
the limit of infinite chemical potentials.
We conclude the paper with a brief summary in Sect. 6.

\section{Model}
\begin{figure}[ht]
\begin{center}
\includegraphics*[scale=0.6]{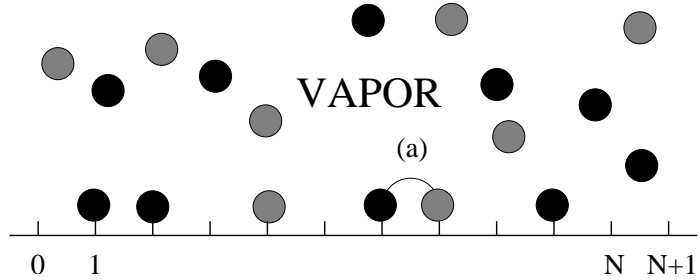}
\caption{
\label{Fig1}
{\small One-dimensional regular lattice of $N$ adsorption sites in contact with a
mixed vapor phase. Black and gray circles denote $A$ and $B$ particles,
respectively. (a) denotes a configuration in which an instantaneous reaction takes
place upon which both particles desorb and the product $AB$ leaves the system.
The sites $i = 0$ and $i = N  + 1$ are always empty.
}
}
\end{center}
\end{figure}

Consider a one-dimensional, regular lattice of $N$ adsorption sites (Fig.~\ref{Fig1}),
which is brought in contact with the mixed vapor phase of two types $A$ and $B$
of hard-core particles without attractive interactions. The difference in chemical
potential between the vapor phase and the adsorbed phase $\mu_A$ and $\mu_B$,
including the binding energy of an occupied site, is maintained constant
(thus $\mu_{A,B} > 0$ corresponds to a preference for adsorption), and the constant
activities are hence defined by
\begin{equation}
z_A = \exp(\beta \mu_A),~~z_B = \exp(\beta \mu_B),
\end{equation}
where $\beta^{-1} = k_B T$ is the thermal energy, $T$ is the temperature,
and $k_B$ is the Boltzmann constant.

The $A$ and $B$ particles can adsorb from the vapor phases onto $vacant$ sites,
and can desorb back to the reservoir. The occupation of the $i$-th site is described
by two Boolean variables $n_i$ and $m_i$, such that
\begin{equation}
n_i = \left\{\begin{array}{ll}
1,     \mbox{ if the $i$-th site is occupied by an $A$ particle,} \nonumber\\
0,     \mbox{ otherwise,}
\end{array}
\right.
\end{equation}
and
\begin{equation}
m_i = \left\{\begin{array}{ll}
1,     \mbox{ if the $i$-th site is occupied by a $B$ particle,} \nonumber\\
0,     \mbox{ otherwise,}
\end{array}
\right.
\end{equation}
and thus the state of the site $i$ is specified by
$(n_i,m_i) \in \{(0,0),(0,1),(1,0),(1,1)\}$. Note that the state $(1,1)$, which
in general is allowed, has a zero probability in the case of hard-core interaction.
(The equivalent description in terms of $\sigma_i = 0,+1,-1$ for empty sites and
sites occupied by $A$ and $B$, respectively, will be discussed in Sect. 4.)
For computational convenience we add two extra ``boundary" sites $i = 0$ and
$i = N  + 1$ at both ends of the chain and stipulate that these sites are always
unoccupied, i.e., $n_0 = n_{N + 1} = m_0 = m_{N+1} = 0$. Furthermore, the system is
locally ``frustrated", i.e., the equilibrium between the chain and the reservoir is
locally perturbed by the ``kinetic" constraint that as soon as an $A$ and a $B$
particle occupy neighboring sites they instantaneously $react$ and leave the chain
(and the $AB$ product leaves the system); therefore, our model corresponds to a
purely adsorption controlled dynamics \cite{krap}.

In equilibrium the partition function of this two-species adsorbate on a chain
of $N$ sites, in which the configurations $\{(n_i,m_i)\}$ with neighboring $A$ and
$B$ or double occupancy of a site are excluded, can be written as:
\begin{equation}
\label{partition}
Z_N = \sum_{\{(n_i,m_i)\}} \prod_{i = 1}^N z_A^{n_i} z_B^{m_i}\Phi_i,
\end{equation}
where $\Phi_i$ is an indicator function with the following properties:
\begin{equation}
\Phi_i = \left\{\begin{array}{llll}
1,     \mbox{ if site $i$ is empty,} \nonumber\\
1,     \mbox{ if site $i$ is occupied by an $A$ particle,
        i.e., $(n_i,m_i)=(1,0)$, while $m_{i\pm1}=0$,} \nonumber\\
1,     \mbox{ if site $i$ is occupied by a $B$ particle,
        i.e., $(n_i,m_i)=(0,1)$, while $n_{i\pm1}=0$,} \nonumber\\
0,     \mbox{ otherwise.}
\end{array}
\right.
\end{equation}
A very simple realization of the indicator function $\Phi_i$ is given by
\begin{equation}
\label{indicator}
\Phi_i = \Big(1 - n_i m_i\Big)
         \Big(1 - n_i m_{i - 1}\Big) \Big(1 - n_i m_{i + 1}\Big)
         \Big(1 - m_i n_{i - 1}\Big) \Big(1 - m_i n_{i + 1}\Big).
\end{equation}
The factor $(1 - n_i m_i)$ ensures that each site $i$ can be occupied by
either $A$ or $B$ particle, or be empty, while the product
$\Big(1 - n_i m_{i - 1}\Big) \Big(1 - n_i m_{i + 1}\Big)$ ensures that if the
site $i$ is occupied by an $A$ the neighboring sites are not occupied by a $B$
(and vice versa for the last two terms).
Note that $\Phi_i$ in Eq.~(\ref{indicator}) is symmetric with respect to
the variables $n_i$ and $m_i$, and therefore the partition function in
Eq.~(\ref{partition}) is a symmetric function of $z_A$ and $z_B$.

The thermodynamics is given by the pressure (in units of the square of the
lattice spacing),
\begin{equation}
\label{pressure}
P \equiv P(T,\mu_A,\mu_B) = \frac{1}{\beta} \lim_{N \to \infty} \frac{\ln Z_N}{N},
\end{equation}
from which all other quantities of interest, such as the mean densities
of the species $A$ and $B$ and the compressibilities can be obtained
straightforwardly by differentiating it suitably with respect to the chemical
potentials $\mu_A$ and $\mu_B$.

Before closing this section we mention two trivial cases:\\
{\bf (a)} If the $A$ and $B$ particles do not react and do not interact, such
that a lattice site can be occupied by an $A$, a $B$, or simultaneously by an
$A$ and a $B$ particle, i.e., $\Phi_i \equiv 1$, the pressure and the mean density
of the $A$ ($B$) particles is given by the classical Langmuir adsorption model
\begin{equation}
\label{lang}
P = \frac{1}{\beta} \ln\left(1 + z_{A, B}\right)
\;\;\; \text{and} \;\;\; n_{A, B} = \frac{z_{A, B}}{1 + z_{A, B}}.
\end{equation}
{\bf (b)} If the $A$ and $B$ particles do not react, but obey the hard-core
exclusion interaction such that a lattice site cannot accommodate more than one
particle (either $A$ or $B$), i.e., $\Phi_i \equiv (1 - n_i m_i)$, one has
\begin{equation}
\label{lang_hc}
P = \frac{1}{\beta} \ln\left(1 + z_A + z_B\right)
\;\;\; \text{and} \;\;\; n_{A,B} = \frac{z_{A,B}}{1 + z_A + z_B}.
\end{equation}
For the following these results will serve as reference expressions, allowing us
to identify that contribution to the thermodynamics of the system which stems from
the ``reaction" part alone.

\section{Calculation of the partition function based on recursion relations}
We define two auxiliary partition functions:
\begin{equation}
\label{partitionA,B}
Z_N^{(A)} =  \left.
             \sum_{\{(n_i,m_i)\}} \prod_{i = 1}^N z_A^{n_i} z_B^{m_i}\Phi_i
             \right|_{n_{N}  = 1}
~~\text{and}~~
Z_N^{(B)} =  \left.
             \sum_{\{(n_i,m_i)\}} \prod_{i = 1}^N z_A^{n_i} z_B^{m_i}\Phi_i
             \right|_{m_{N}  = 1},
\end{equation}
which are the partition functions of the chain under the constraint that the
site $i = N$ is occupied by an $A$ particle (i.e., $n_N = 1$) or a $B$ particle
(i.e., $m_N = 1$), respectively. Summing over the occupation variables of the
site $i = N$ yields the following relation for the unconstrained partition function
(Eq.~(\ref{partition})):
\begin{equation}
\label{rec}
Z_N = Z_{N -1} + Z_N^{(A)} + Z_N^{(B)},~~\textrm{for}~N \geq 2.
\end{equation}
Summing over the occupation variables of the site $i = N-1$ for
$n_N = 1$ or $m_N = 1$, respectively, one finds that the constrained
partition functions (Eq.~(\ref{partitionA,B})) satisfy
\begin{equation}
\label{A}
Z_N^{(C)} = z_{C} Z_{N -2} + z_{C} Z_{N - 1}^{(C)},
~~\textrm{for}~N \geq 3,~C = A,\;B.
\end{equation}
Equations (\ref{rec}) and (\ref{A}), augmented with the obvious initial
conditions
\begin{eqnarray}
\label{initial}
&&Z_1 = 1 + z_A + z_B,\nonumber\\
&&Z_1^{(A)} = z_A, \;\;\; Z_2^{(A)} = z_A + z_A^2, \nonumber\\
&&Z_1^{(B)} = z_B, \;\;\; Z_2^{(B)} = z_B + z_B^2,
\end{eqnarray}
represent a closed system of coupled recursion relations determining $Z_N$ for
arbitrary $N$, $z_A$, and $z_B$.

The solution of Eqs.~(\ref{rec})-(\ref{initial}) can be found using the standard
generating function technique. Let
\begin{equation}
\label{par}
{\cal Z}_t = \sum_{N = 1}^{\infty} Z_N t^N,~~~
{\cal Z}^{(C)}_t = \sum_{N = 1}^{\infty} Z_N^{(C)} t^N,~C = A,\;B,
\end{equation}
denote such generating functions.
Multiplying Eqs.~(\ref{rec}) and (\ref{A}) by $t^N$, summing over $N$,
and using the expressions in Eq.~(\ref{initial}) for the corresponding
coefficients of the terms in $t$ and $t^2$, we find that ${\cal Z}_t$,
${\cal Z}^{(A)}_t$, and ${\cal Z}^{(B)}_t$ obey the following equations:
\begin{eqnarray}
&&\Big( 1 - t\Big) {\cal Z}_t =  t + {\cal Z}^{(A)}_t  + {\cal Z}^{(B)}_t, \nonumber\\
&&\Big( 1 - z_A t\Big) {\cal Z}^{(A)}_t =
z_A t^2  {\cal Z}_t + z_A t \Big(1 + t\Big), \nonumber\\
&&\Big( 1 - z_B t\Big) {\cal Z}^{(B)}_t =
z_B t^2  {\cal Z}_t + z_B t \Big(1 + t\Big).
\end{eqnarray}
Consequently, ${\cal Z}_t$ is a rational function of $t$ and is given explicitly by
\begin{equation}
\label{par_pol}
{\cal Z}_t = \frac{t {\cal L}_1(t)}{{\cal L}_2(t)},
\end{equation}
where
\begin{equation}
{\cal L}_1(t) =  \frac{1 + z_A + z_B}{z_A z_B} - 2  t - t^2,
~~~
{\cal L}_2(t) = \frac{1}{z_A z_B} - \frac{1 + z_A + z_B}{z_A z_B} t + t^2 +  t^3.
\end{equation}
Let $t_1$, $t_2$ and $t_3$ be the roots of ${\cal L}_2(t)$ so that
${\cal L}_2(t) = (t-t_1)(t-t_2)(t-t_3)$. Expressing  Eq.~(\ref{par_pol}) in terms of
elementary fractions and expanding the emerging factors $1/(t_j - t)$, $j = 1,2,3$,
into Taylor series in powers of $t/t_j$, Eq.~(\ref{par_pol}) can be cast into the form
\begin{equation}
\label{part}
{\cal Z}_t = \sum_{N = 1}^{\infty} \left\{\alpha^{(1)} \Big(\frac{t}{t_1}\Big)^N
+  \alpha^{(2)} \Big(\frac{t}{t_2}\Big)^N + \alpha^{(3)} \Big(\frac{t}{t_3}\Big)^N\right\},
\end{equation}
where the coefficients $\alpha^{(j)}$ are given explicitly by
\begin{equation}
\alpha^{(1)} = \frac{t_1+t_2 t_3}{(t_1 - t_2) (t_1 - t_3)},~~
\alpha^{(2)} = \frac{t_2+t_1 t_3}{(t_2 - t_1) (t_2 - t_3)},~~
\alpha^{(3)} = \frac{t_3+t_1 t_2}{(t_3 - t_1) (t_3 - t_2)}.
\end{equation}

Equations (\ref{par}) and (\ref{part}) imply that the partition function of a chain
with $N$ sites (Eq.~(\ref{partition})) is given explicitly by
\begin{equation}
Z_N = \frac{\alpha^{(1)}}{ t_1^N}
+  \frac{\alpha^{(2)}}{ t_2^N} +
\frac{\alpha^{(3)}}{t_3^N},
\end{equation}
and thus its behavior is completely determined by the properties of the roots of the
cubic polynomial ${\cal L}_2(t)$. Defining
\begin{equation}
\label{q}
q = \frac{3 + 3 z_A + 3 z_B + z_A z_B}{54 z_A z_B},~~
r = \frac{36 + 9 z_A + 9 z_B + 2 z_A z_B}{9 z_A z_B},
\textrm{~~and~}
X = \frac{r}{q^{3/2}},
\end{equation}
one can show that $Q = r^2 - q^3 < 0$ so that $0 < X < 1$ for all $z_{A,B}>0$, which
implies that all three roots of ${\cal L}_2(t)$ are real and given by \cite{abr}
\begin{equation}
\label{t1}
t_{1,3} = \pm 2 \sqrt{q} \cos\left(\pm\frac{\pi}{6} +
           \frac{1}{3} \arcsin(X)\right) - \frac{1}{3},
\end{equation}
\begin{equation}
\label{t2}
t_2 = 2 \sqrt{q} \sin\left(\frac{1}{3} \arcsin(X)\right) - \frac{1}{3}.
\end{equation}

The roots of ${\cal L}_2(t)$ satisfy
$\displaystyle{t_1 t_2 t_3 = -\frac{1}{z_A z_B} < 0}$,
$\displaystyle{t_1 t_2+t_1 t_3+t_2 t_3 = - \frac{(1 + z_A + z_B)}{z_A z_B}} < 0$, and
since all three are real, it follows that one root is negative and the other two
are positive. Using $0 < \arcsin(X) < \pi/2$, it is easy to prove that for any $z_A$
and $z_B$ one has $t_1 > t_2 > t_3$ (thus $t_3$ is the negative solution) and
$|t_3|> t_1$. Consequently, due to Eq.~(\ref{pressure}), in the thermodynamic limit
$N \to \infty$ the pressure is given by
\begin{equation}
\label{k}
P = - \frac{1}{\beta}
\ln\left(2 \sqrt{q} \sin\left(\frac{1}{3} \arcsin(X)\right) - \frac{1}{3} \right).
\end{equation}

We note that here three different roots appear since the model under study includes
effectively three-site interactions, as shown by the indicator function $\Phi_i$
in Eq.~(\ref{indicator}). The fact that as function of the chemical potentials neither
pair of roots does intersect for any $z_A$ and $z_B$ implies, as expected, that the
one-dimensional model considered here does not exhibit a phase transition. More
complicated models involving particles which require more than one empty lattice site
for their adsorption, such as, for instance, the original ZGB model in which $A$ species
are dimers \cite{zgb}, can be treated within essentially the same approach as presented
here. Such models would include effective four-site interactions (or more), and thus
four or more different roots will emerge.

\section{Connection with classical spin S = 1 models}
Since, as we have noted, in the present model the hard-core interaction excludes the
state $(n_i,m_i) = (1,1)$ for a site, there is a natural connection between the present
model and the three-state lattice gas model \cite{3state} or, equivalently, a
special Blume-Emery-Griffiths (BEG) spin $S = 1$ model \cite{BEG}.

The mapping to the BEG model is accomplished as follows. We assign to each site of the
one-dimensional lattice a three-state variable $\sigma_i,~i=1,\dots,N$, such that
\begin{equation}
\sigma_i =
\begin{cases}
+1, & ~~\textrm{if the $i$-th site is occupied by an $A$ particle,} \cr
-1, & ~~\textrm{if the $i$-th site is occupied by a $B$ particle,} \cr
~~0,  & ~~\textrm{if the $i$-th site is empty.}\cr
\end{cases}
\label{spin}
\end{equation}
In terms of $\sigma_i$ the occupation numbers $n_i$ and $m_i$ may be rewritten as
\begin{equation}
n_i=(\sigma_i+\sigma_i^2)/2~,~~~ m_i=(-\sigma_i+\sigma_i^2)/2.
\label{oc_num}
\end{equation}
Periodic boundary conditions imply $\sigma_{N+1} \equiv \sigma_1$.
Defining the nearest-neighbor (NN) coupling as (the limits corresponding to our
particular model are indicated in parenthesis)
\begin{equation}
J_{i,j}=
\begin{cases}
-E_1~(\to 0), & ~~\textrm{for A-A neighbors, }\cr
-E_2~(\to 0), & ~~\textrm{for B-B neighbors, }\cr
+E_3~(\to \infty), & ~~\textrm{for A-B or B-A neighbors, }\cr
~~~~0, & ~~\textrm{otherwise, } \cr
\end{cases}
\label{coupling}
\end{equation}
one can write the Hamiltonian of this system as
\begin{equation}
\mathcal{H}=\sum_{<ij>}\left[n_i n_j(-E_1)+m_i m_j(-E_2)+(n_i m_j + n_j m_i) E_3\right]
-\sum_{i=1}^{N}(\mu_A n_i+\mu_B m_i)
\label{hamiltonian_ij}
\end{equation}
where $\displaystyle{\sum_{<ij>}}$ means summation over all pairs of NN sites with
each pair included only once. Replacing $n_i$, $m_i$ with the corresponding expressions
(Eq.~(\ref{oc_num})) and collecting the terms, the Hamiltonian above may be rewritten as
\begin{eqnarray}
&&\mathcal{H}=-\frac{E_1+E_2+2 E_3}{4}\sum_{<ij>}\sigma_i \sigma_j
     -\frac{E_1+E_2-2 E_3}{4}\sum_{<ij>}\sigma_i^2 \sigma_j^2
     -\frac{E_1-E_2}{4}\sum_{<ij>}\left(\sigma_i \sigma_j^2+\sigma_j \sigma_i^2\right)
     \nonumber\\
&& ~~~~~~-\frac{\mu_A -\mu_B}{2}\sum_{i=1}^{N} \sigma_i
     -\frac{\mu_A +\mu_B}{2}\sum_{i=1}^{N} \sigma_i^2,
\label{H_BEG}
\end{eqnarray}
i.e., takes the form of a Hamiltonian for a spin $S=1$ model \cite{furman}
with the parameters\hfill\\
$$
J=\frac{E_1+E_2+2 E_3}{4},~K=\frac{E_1+E_2-2 E_3}{4},~C=\frac{E_1-E_2}{4},~
H=\frac{\mu_A -\mu_B}{2},~\textrm{and } \Delta=-\frac{\mu_A +\mu_B}{2}.
$$

Therefore, our model ($E_1=E_2=0$, $E_3 \to \infty$) is equivalent with a BEG model
characterized by a rather unusual set of interaction parameters: $C=0$,
a bilinear exchange constant $J = E_3/2 \to \infty$, and a biquadratic exchange constant
$K = -E_3/2 \to -\infty$ such that $J/K = -1$.

The description in terms of the spin variables $\sigma_i$ opens an alternative way
of computing the partition function in Eq.~(\ref{partition}) by using the standard
transfer-matrix method \cite{Baxter}.
Although it is possible to set up the transfer matrix using directly the Hamiltonian
given in Eq.~(\ref{H_BEG}), it turns out to be more convenient to define local
fields $\mu(\sigma_i)$ as
\begin{equation}
\mu(\sigma_i) =
\begin{cases}
-\mu_A,  &   \mbox{ if $\sigma_i = 1$,} \cr
+ \mu_B,  &   \mbox{ if $\sigma_i = - 1$,} \cr
0,  &   \mbox{ if $\sigma_i = 0$,} \cr
\end{cases}
\end{equation}
such that the partition function for the system is written as
\begin{equation}
\label{i}
Z_N^{(BEG)} = \sum_{\{\sigma_i\}}
\exp\left[ \sum_{i=1}^{N} \left(- \beta J_{i,i+1} \sigma_i \sigma_{i+1} -
\beta \mu(\sigma_i) \sigma_i\right)\right],
\end{equation}
or, equivalently, as the trace of the product of transfer matrices
\begin{equation}
Z_N^{(BEG)} = {\rm Tr} \prod_{i = 1}^N V_{i,i+1},
\end{equation}
where the transfer matrix $V_{i,i+1}$ is given by
\begin{equation}
V_{i,i+1} = \exp\left[- \beta J_{i,i+1} \sigma_i \sigma_{i+1} -
\beta \Big(\mu(\sigma_i)\sigma_i + \mu(\sigma_{i+1}) \sigma_{i+1}\Big)/2\right].
\end{equation}
In the thermodynamic limit the expressions for the pressure given by the partition
functions in Eqs.~(\ref{partition}) and (\ref{i}) become identical if $E_1 = E_2 = 0$,
and $E_3 \to \infty$. For these values of the parameters, the transfer matrix
$V_{i,i+1}$ becomes
\begin{equation}
\label{AA}
{\rm V}_0 = {\rm V}_{i,i+1} =
\begin{pmatrix}
z_A& \sqrt{z_A}& 0& \\
\sqrt{z_A}& 1    & \sqrt{z_B}& \\
0& \sqrt{z_B} & z_B&
\end{pmatrix}.
\end{equation}
The eigenvalues $\lambda_j = 1/t_j$ of ${\rm V}_0$ are given by
${\rm det}\Big(V_0 - \lambda I\Big) = 0$, which are the inverse of the zeroes
of ${\cal L}_2(t) = 0$.

\section{Mean density and compressibility}
\subsection{Thermodynamics}
Since the model and the partition sum are symmetric with respect to interchanging $A$
and $B$ particles, it is sufficient to analyze the behavior of only one of the two
densities, e.g., $n_A$; the corresponding results for $n_B$, and similarly for the
compressibilities, follow by simply switching the indices $A$ and $B$.

\begin{figure}[!htb]
\begin{center}
\begin{minipage}[c]{.5\textwidth}
\includegraphics[width=.9\textwidth]{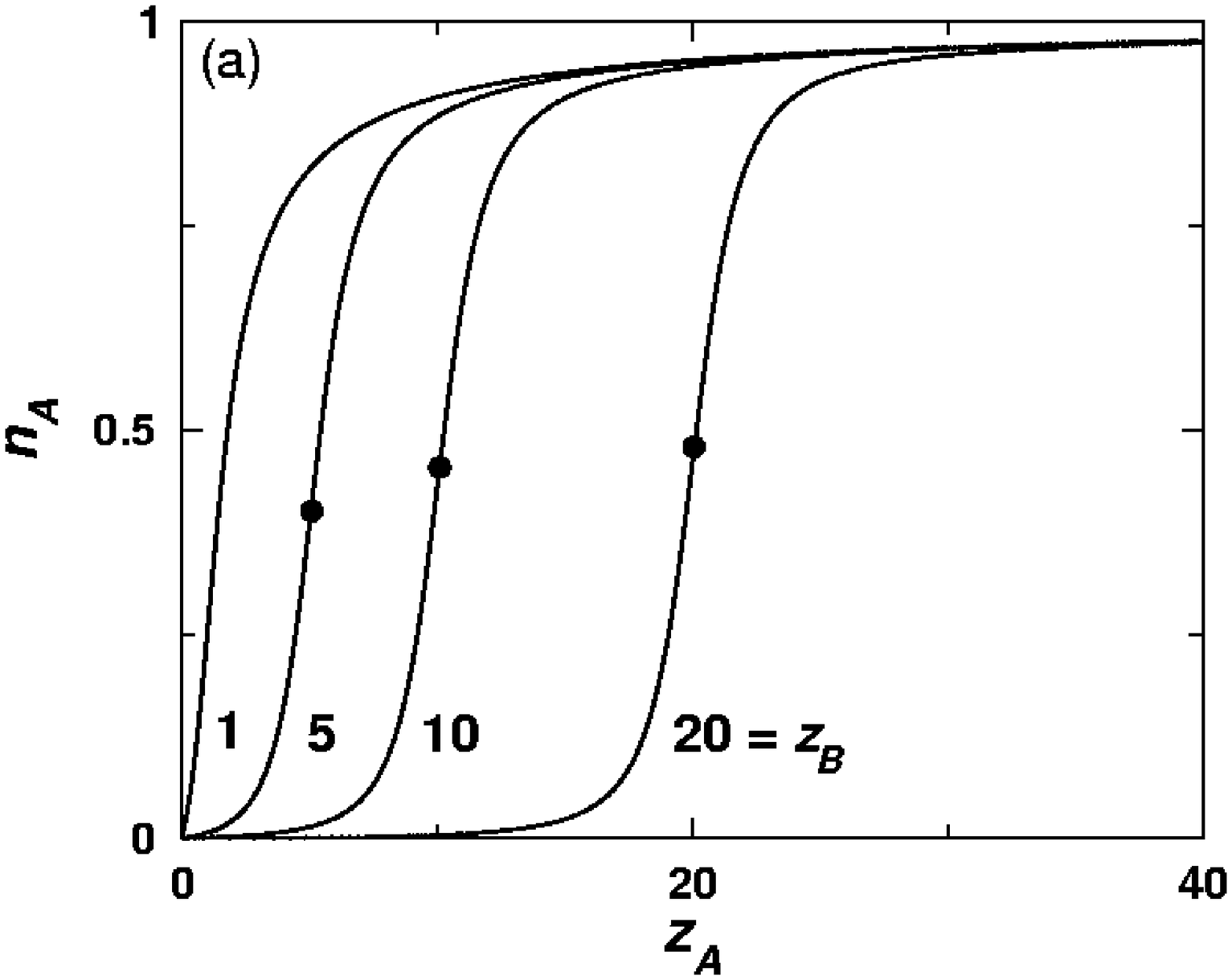}%
\hspace*{-1.2in}%
\raisebox{.22\textwidth}{
\includegraphics[width=.33\textwidth]{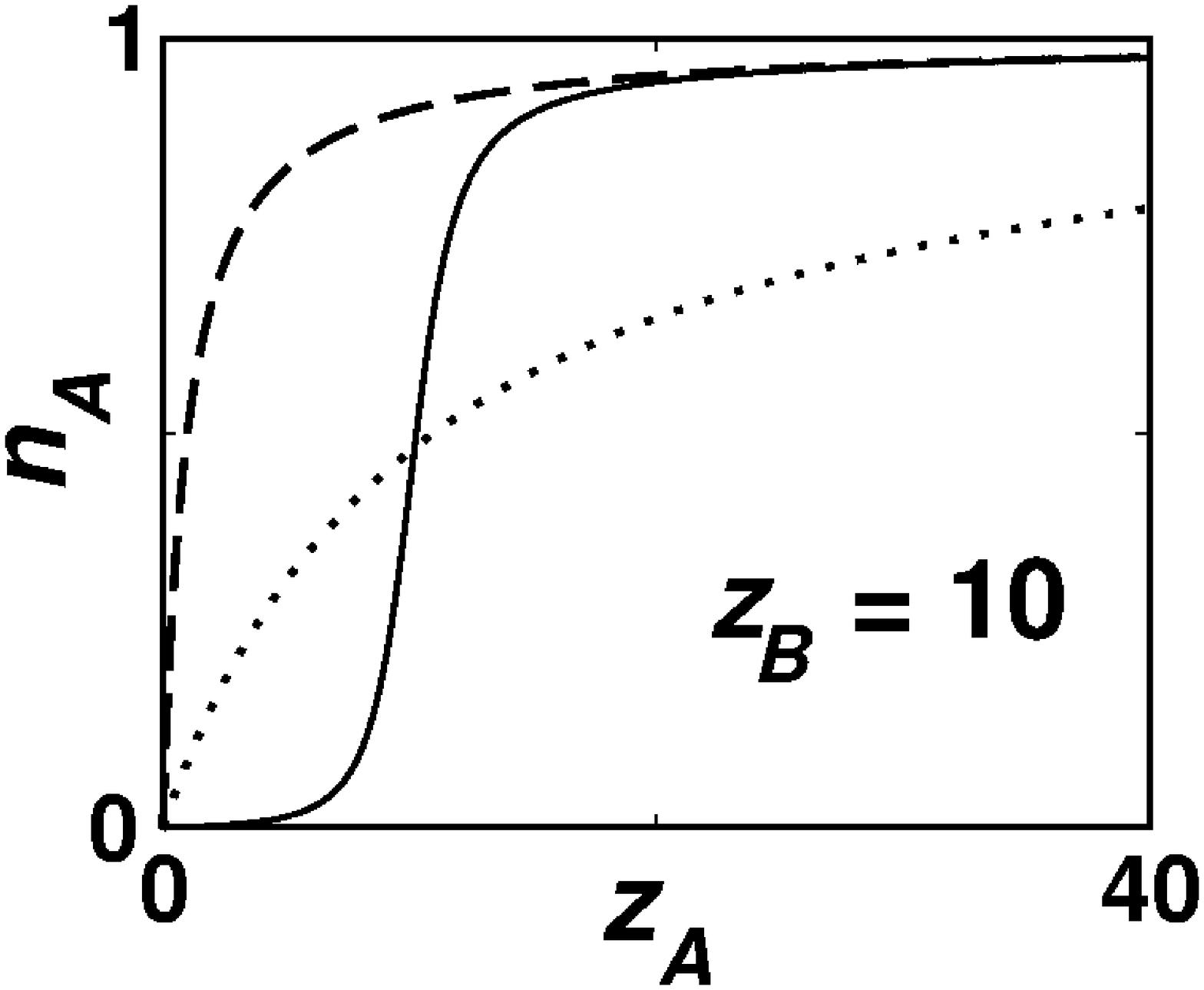}
}
\end{minipage}%
\begin{minipage}[c]{.5\textwidth}
\includegraphics[width=.9\textwidth]{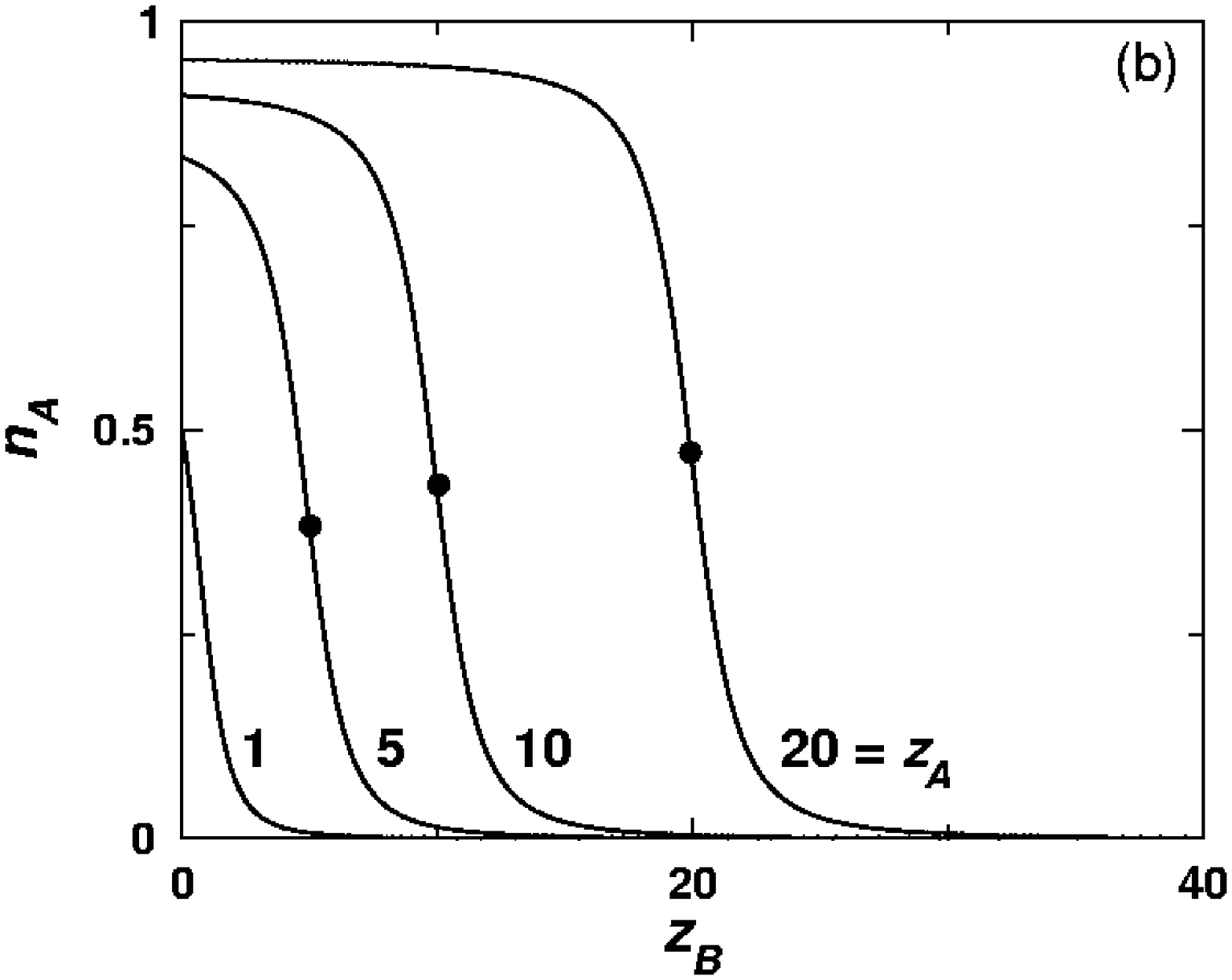}%
\hspace*{-1.2in}%
\raisebox{.27\textwidth}{
\includegraphics[width=.33\textwidth]{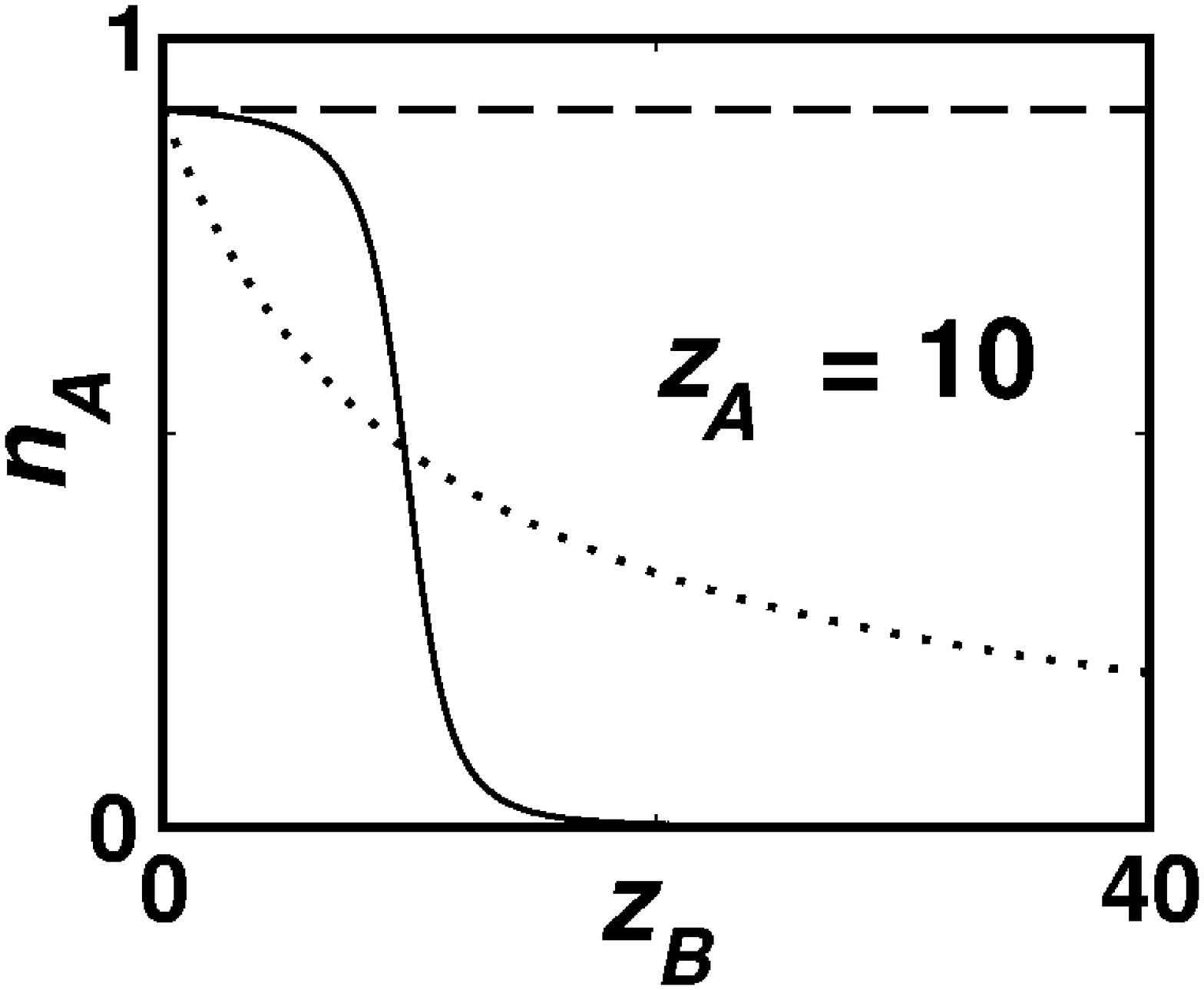}
}
\end{minipage}
\caption{
\label{Fig2}
{\small Mean density $n_A$ as a function of $z_A$ (a)
and $z_B$ (b), respectively. From left to right, the curves correspond to (a)
$z_B = 1,~5,~10$, and $20$ and to (b) $z_A = 1,~5,~10$, and $20$, respectively.
The dots indicate the turning points $(z_A^*,n_A^*)$ and $(z_B^*,n_A^*)$.
For comparison, the insets show results for $z_B=10$ (a) and for $z_A=10$ (b),
respectively, together with corresponding results for the case of no $AB$ reaction
(classical Langmuir, Eq.~(\ref{lang}), dashed line) and for the approximation
given by Eq.~(\ref{lang_hc}) (hard-core exclusion only, dotted line).
}
}
\end{center}
\end{figure}

Using Eq.~(\ref{k}), the mean density $n_A$ and the compressibility $\varkappa_A$
of the phase formed by the adsorbed $A$ particles are given as
\begin{equation}
\label{density}
n_A = \beta z_A \frac{\partial P}{\partial z_A},~~~~
\varkappa_A = \frac{z_A}{n_A^2} \frac{\partial n_A}{\partial z_A}.
\end{equation}
The behavior of $n_A$ as a function of $z_A$ and $z_B$, respectively, is shown in
Fig.~(\ref{Fig2}) for several values of $z_B$ and $z_A$, respectively.

\begin{figure}[!htb]
\begin{center}
\begin{minipage}[c]{.5\textwidth}
\includegraphics[width=.9\textwidth]{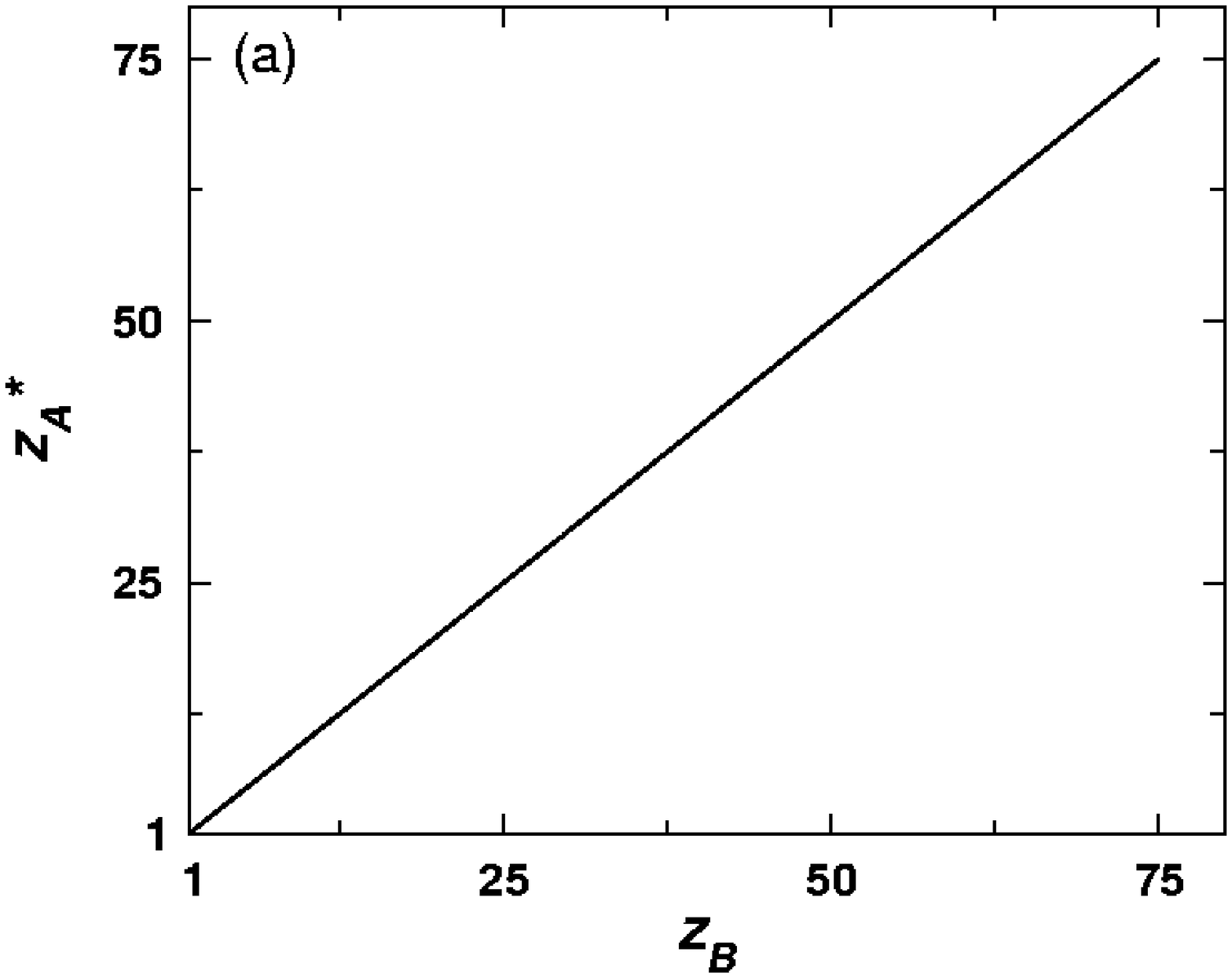}
\end{minipage}%
\begin{minipage}[c]{.5\textwidth}
\includegraphics[width=.9\textwidth]{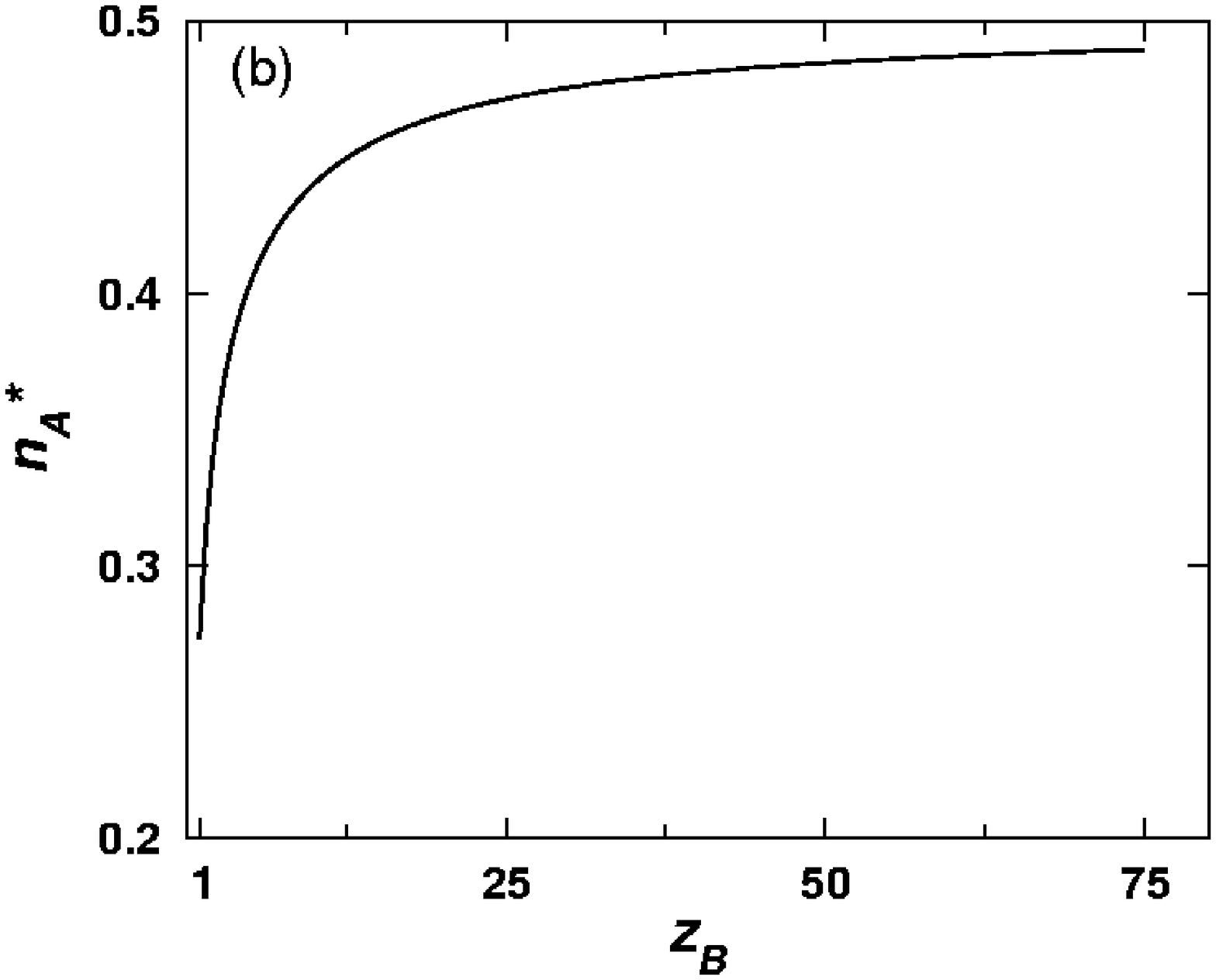}
\end{minipage}
\caption{\label{Fig2add} {\small The coordinates $(z_A^*$, $n_A^*)$ of the turning point
of the curves $n_A(z_A)$ as functions of $z_B$.
}
}
\end{center}
\end{figure}

As can be seen in Fig.~\ref{Fig2}, the density $n_A$ is very small for $z_A < z_A^*$
(where $(z_A^*,n_A^*)$ is the turning point of $n_A(z_A)$, occuring for $z_B > 1$),
raises sharply to large values for $z_A > z_A^*$ and, independent of the {\it fixed}
value for $z_B$, approaches asymptotically $n_A(z_A \to \infty;z_B) \to 1$. The functions
$z_A^*(z_B)$ and $n_A^*(z_B)$ depicted in Fig.~\ref{Fig2add} show that the value $z_A^*$
is an almost linear function of $z_B$, satisfying $z_A^*(z_B) \gtrapprox z_B$, while
the density $n_A^* = n_A(z_A^*(z_B),z_B)$ raises sharply with increasing $z_B$ and
saturates at the asymptotic value $n_A^*(z_B \to \infty) \to 0.5$.
The width $\Delta z_A$ of the sharp increase in $n_A$ depends weakly on $z_B$ and
at fixed $z_B$ the function $n_A(z_A;z_B)$ converges slowly toward the step function
$\Theta(z_A-z_B)$ with increasing $z_B$. In this limit the system exhibits two generic
states: an $A$-depleted one ($n_A \ll 1$, for $z_A < z_A^*$) and an $A$-saturated
one ($n_A \lesssim 1$, for $z_A > z_A^*$), for which the density $n_A$ depends only
very weakly on $z_A$, separated by a narrow transition region centered at
$z_A = z_A^*$. Similar conclusions can be drawn regarding the behavior of the mean
density $n_A$ as a function of $z_B$.
Comparing these results with the ones corresponding to Langmuir adsorption
(Eqs.~(\ref{lang}) and~(\ref{lang_hc})) it is evident that in general the
reaction $AB$ leads to a {\it qualitatively} different behavior of the mean
density $n_A$ (see the insets in Fig.~\ref{Fig2}). We note that, as expected, at
fixed $z_A$ (inset Fig.~\ref{Fig2}(b)) the behavior for $z_B \ll 1$ is well described
by the classical Langmuir non-interacting particles model but, rather
counter-intuitively, the behavior at large $z_A$ and fixed $z_B$ (inset
Fig~\ref{Fig2}(a)) is also similar to that of a non-interacting case and not similar
to the corresponding case with hard-core interaction. These findings will be addressed
in the discussion of the asymptotic behavior below.

\begin{figure}[!htb]
\begin{center}
\begin{minipage}[c]{.5\textwidth}
\includegraphics[width=.9\textwidth]{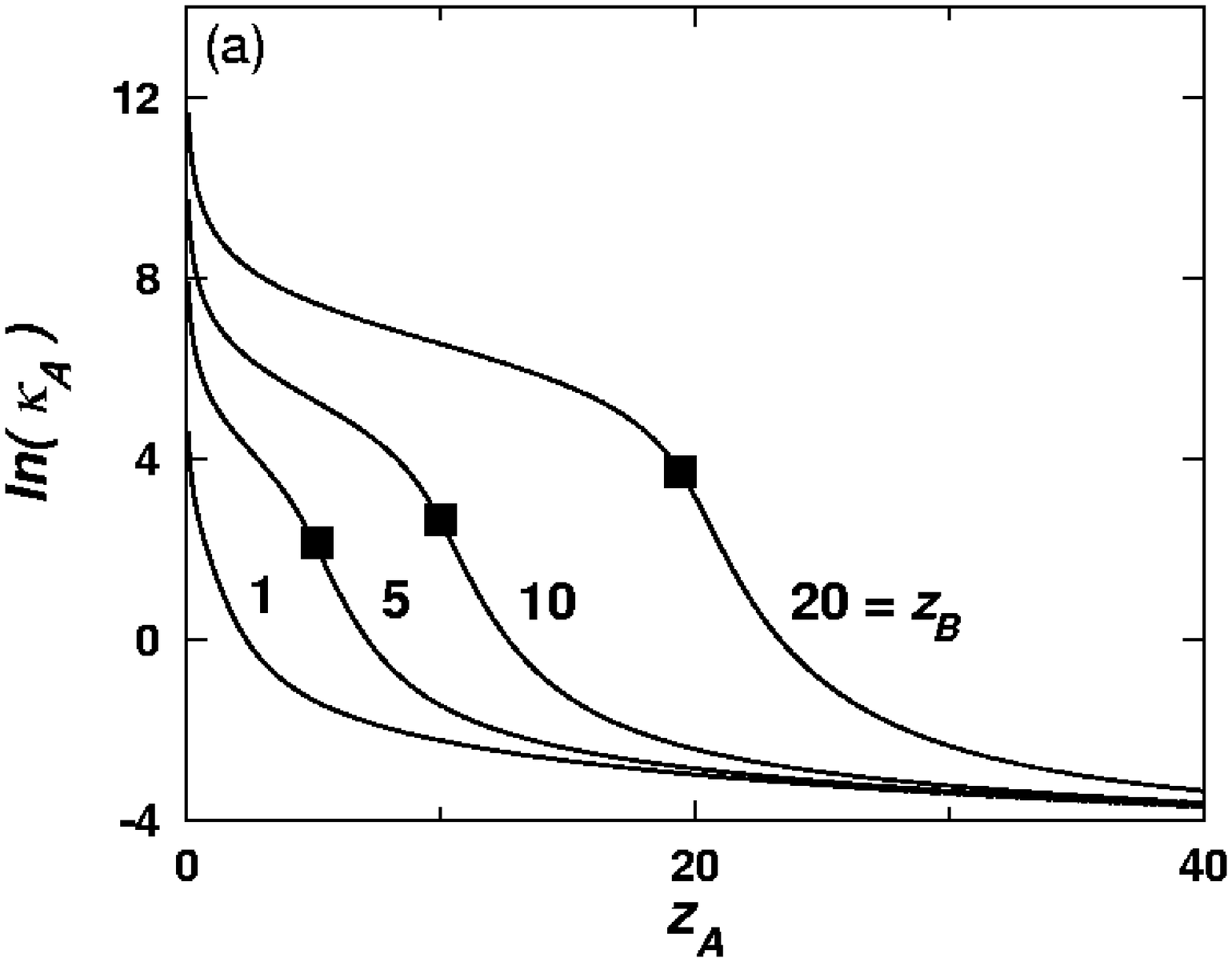}
\end{minipage}%
\begin{minipage}[c]{.5\textwidth}
\includegraphics[width=.9\textwidth]{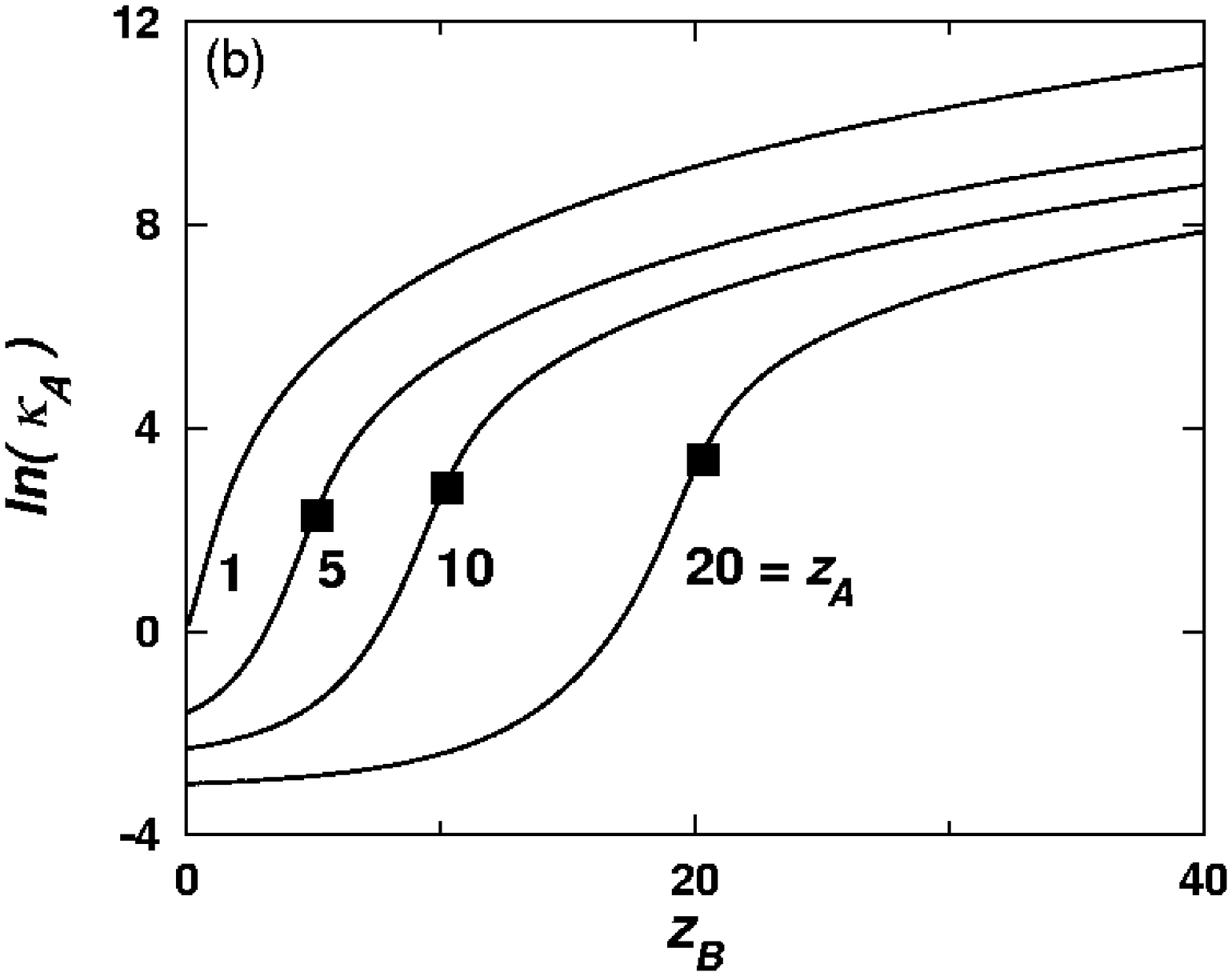}
\end{minipage}
\caption{\label{Fig3} {\small Logarithm of compressibility $\varkappa_A$
as a function of $z_A$ (a) and $z_B$ (b), respectively. From left to right,
the curves correspond to (a) $z_B = 1,~5,~10$, and $20$ and (b) $z_A = 1,~5,~10$,
and $20$, respectively. The squares denote the position of the turning points of
$n_A(z_A)$ (Fig. 2(a)) and  $n_A(z_B)$ (Fig. 2(b)), respectively.
}
}
\end{center}
\end{figure}

In Fig.~\ref{Fig3} we show the compressibility $\varkappa_A$ of the phase formed by
the adsorbed $A$ particles as a function of the activities $z_A$ and $z_B$,
respectively, for the same values as in Fig.~\ref{Fig2}.
The behavior of the compressibility is consistent with that of the mean density,
i.e., for example Fig.~\ref{Fig2}(a) shows that the low $A$-density phase
at $z_A < z_A^*$ has a large and slowly decreasing compressibility, while the high
$A$-density phase at $z_A > z_A^*$ has a very small compressibility.

As indicated by Eqs.~(\ref{k}) and (\ref{density}), the functions $n_A (z_A,z_B)$
and $\varkappa_A (z_A,z_B)$ exhibit complicated dependences on $z_A,~z_B$, and thus
an analytical study of the behavior of the system for all the values of the
parameters is difficult. However, as we have mentioned, for fixed values
of one of the activities, for example $z_B$, there are three particular cases
as a function of the other one, i.e., $z_A$: $z_A \ll z_A^*$, $z_A \gg z_A^*$,
and $z_A \simeq z_A^*$. Below we derive and discuss the asymptotic behavior
in these limits, starting from corresponding expansions of $n_A (z_A,z_B)$
or, if necessary, of $t_2 (z_A,z_B)$ (Eq.~(\ref{t2})).

\subsection{Asymptotic expansions}
We first consider the case when either one or both activities are small, i.e.,
the corresponding vapor pressure is low, the temperature is high, or the barrier
against desorption is small.
In the case when $z_A \ll 1$, while $z_B$ is fixed, we find from Eqs.~(\ref{k})
and (\ref{density}):
\begin{equation}
\label{nn}
n_A = \frac{1}{\left(1 +z_B\right)^3} z_A +
\frac{\left(2 z_B^2 + 6 z_B - 1\right)}{\left(1 +z_B\right)^6} z_A^2 +
{\cal O}\left(z_A^3\right),
\end{equation}
and thus the compressibility is
\begin{equation}
\varkappa_A = \frac{\left(1 +z_B\right)^3}{z_A} -
\frac{\left(z_B^3 + 6 z_B^2 + 11 z_B + 9\right) z_B}{\left(1 +z_B\right)^3} z_A +
{\cal O}\left(z_A^2\right).
\end{equation}
We note that in Eq.~(\ref{nn}) the coefficient of the term linear in $z_A$ is smaller
by a factor $\left(1 +z_B\right)^{-3}$ than the corresponding one in the case of
non-interacting Langmuir adsorption (Eq.~(\ref{lang})) and by a factor
$\left(1 +z_B\right)^{-2}$ compared with hard-core Langmuir adsorption
(Eq.~(\ref{lang_hc})). Consequently, the compressibility is increased by similar
factors. This reduction of $n_A$ is significant for large $z_B$ and shows that in
the presence of many adsorbed $B$ particles most of the adsorbed $A$ particles do
react with $B$ and leave the chain. In the opposite case $z_B \ll 1$ the first-order
term is the same as for the non-interacting Langmuir adsorption (Eq.~(\ref{lang}))
since this limit corresponds to very low densities of both $A$ and $B$ particles and,
consequently, very unlikely reaction events.

Second, within the limit $z_B \ll 1$, while $z_A$ is fixed, we obtain
\begin{equation}
n_A = \frac{z_A}{1 +z_A}  - \frac{3 z_A}{\left(1 +z_A\right)^4} z_B -
\frac{\left(4 z_A^2 + 13 z_A - 6\right) z_A}{\left(1 + z_A\right)^7} z_B^2 +
{\cal O}\left(z_B^3\right)
\end{equation}
and
\begin{equation}
\varkappa_A = \frac{1}{z_A}  +
\frac{3 \left(3 z_A + 1\right)}{z_A \left(1 +z_A\right)^3} z_B +
{\cal O}\left(z_B^2\right).
\end{equation}
As expected, the first term in the series for both $n_A$ and $\varkappa_A$ is the
trivial Langmuir adsorption model result (Eq.~(\ref{lang}); see also the inset in
Fig.~\ref{Fig2}(b)), because the very low density of adsorbed $B$ particles leads
to a very small probability for a reaction $AB$.

Third, for $z_A = z_B = z \ll 1$ we find
\begin{equation}
n_A = n_B = z - 4 z^2 + 19 z^3 + {\cal O}\left(z^4\right)
\end{equation}
and
\begin{equation}
\varkappa_A = \frac{1}{z} + 3 + 3 z - 8 z^2 + {\cal O}\left(z^3\right),
\end{equation}
which shows that at small activities the value of the density $n_A$ (or $n_B$)
at the crossover $z_A = z_B$ increases linearly with the activity.

Next we turn to the case when either one or both of activities are large, which
can be realized in systems with low temperature, suppressed desorption, or at
high pressures of the corresponding vapor phase. Most of the previous theoretical
work has been focused on this limit \cite{dic}. Since in this case the analysis is
somewhat more complicated compared to the situations with small activities, it is
advantageous to start from the asymptotic behavior of $t_2$ given in Eq.~(\ref{t2}).

First, we consider the case $z_A \gg 1$ and $z_B$ fixed. Using the identity
\begin{equation}
\sin\left(\frac{1}{3} \arcsin\left(
\frac{\sqrt{z_B} (9  + 2 z_B)}{2 (3 + z_B)^{3/2}}
\right)\right) = \frac{1}{2} \sqrt{\frac{z_B}{3 + z_B}}~,
\end{equation}
one finds for the Laurent series of $t_2$
\begin{equation}
t_2 = \frac{1}{z_A} - \frac{1}{z_A^2} +
\frac{1}{z_A^3} - \frac{\left(1+z_B\right)}{z_A^4} +
{\cal O}\left(\frac{1}{z_A^5}\right).
\end{equation}
This implies for the pressure
\begin{equation}
\beta P = \ln\left(z_A\right) + \frac{1}{z_A} - \frac{1}{2 \, z_A^2} +
\frac{\left(1 + 3 z_B\right)}{3 \, z_A^3} + {\cal O}\left(\frac{1}{z_A^4}\right)
\end{equation}
and, hence, for the mean density of the $A$ particles
\begin{equation}
\label{denn}
n_A = 1 - \frac{1}{z_A} +  \frac{1}{z_A^2} -
\frac{\left(1 + 3 z_B\right)}{z_A^3} + {\cal O}\left(\frac{1}{z_A^4}\right).
\end{equation}
Therefore, in this limit the dependence on $z_B$ appears only in the third-order term
and thus it is very weak. This explains the confluence of the density curves $n_A$
in Fig.~\ref{Fig2}(a) in the range of large $z_A$. Here we also note that regardless
of the value of $z_B$ (provided that $z_B \ll z_A$) the mean density of the $A$
particles tends to unity, i.e., the one-dimensional chain becomes saturated with
$A$ particles.
Moreover, the first three terms in the expansion in Eq.~(\ref{denn}) coincide with the
corresponding terms in the expansion of the mean particle density for large $z_A$ in
the non-interacting Langmuir adsorption model (Eq.~(\ref{lang})). However, only the
first term agrees with the similar expansion of the hard-core Langmuir adsorption
(Eq.~(\ref{lang_hc})). This confirms the numerical results shown in the inset in
Fig.~\ref{Fig2}(a) and suggests that the reaction term exactly cancels the
contribution of the hard-core interaction up to the third order in $1/z_A$.
Intuitively, this can be understood by noting that for $z_A \gg z_B$ the system
tends to a state with a very low density $n_B$, and thus the hard-core constraint that
a site cannot be occupied simultaneously by an $A$ and a $B$ particle is effectively
irrelevant, but the constraint that a site cannot be occupied by two $A$ particles
becomes very important.
In this limit we obtain for the compressibility
\begin{equation}
\varkappa_A =  \frac{1}{z_A} + \frac{9 z_B}{z_A^3} +
{\cal O}\left(\frac{1}{z_A^4}\right).
\end{equation}
This confirms that in leading order the adsorption is given by the Langmuir-type
result for the non-interacting adsorption model with corrections occurring only in
third order.

In the opposite limit $z_B \gg 1$ and $z_A$ fixed, similar calculations lead to
\begin{equation}
t_2 = \frac{1}{z_B} - \frac{1}{z_B^2} + \frac{1}{z_B^3}
- \frac{\left(1+z_A\right)}{z_B^4} + \frac{\left(1 + 4 z_A - z_A^2\right)}{z_B^5}
+ {\cal O}\left(\frac{1}{z_B^6}\right),
\end{equation}
which implies
\begin{equation}
n_A = \frac{z_A}{z_B^3} - \frac{\left(3 - 2 z_A\right) z_A}{z_B^4} +
\frac{\left(6 - 6 z_A + 3 z_A^2\right) z_A}{z_B^5} +
{\cal O}\left(\frac{1}{z_B^6}\right),
\end{equation}
and, respectively,
\begin{equation}
\varkappa_A = \frac{z_B^3}{z_A} + \frac{3 z_B^2}{z_A} + {\cal O}\left(z_B\right).
\end{equation}
Thus $n_A$ exhibits a very fast power-law decay implying a strong divergence of the
compressibility $\varkappa_A$ for increasing $z_B$.

Lastly, for $z_A = z_B = z \gg 1$ we find
\begin{equation}
n_A (= n_B) = \frac{1}{2} - \frac{1}{z} + \frac{4}{z^2} +
{\cal O}\left(\frac{1}{z^3}\right),~~~~
\label{na_half}
\varkappa_A = \frac{4}{z} - \frac{16}{z^2} + {\cal O}\left(\frac{1}{z^3}\right).
\end{equation}

\subsection{Monte Carlo simulations}
\begin{figure}[!htb]
\begin{center}
\begin{minipage}[c]{.5\textwidth}
\includegraphics[width=.9\textwidth]{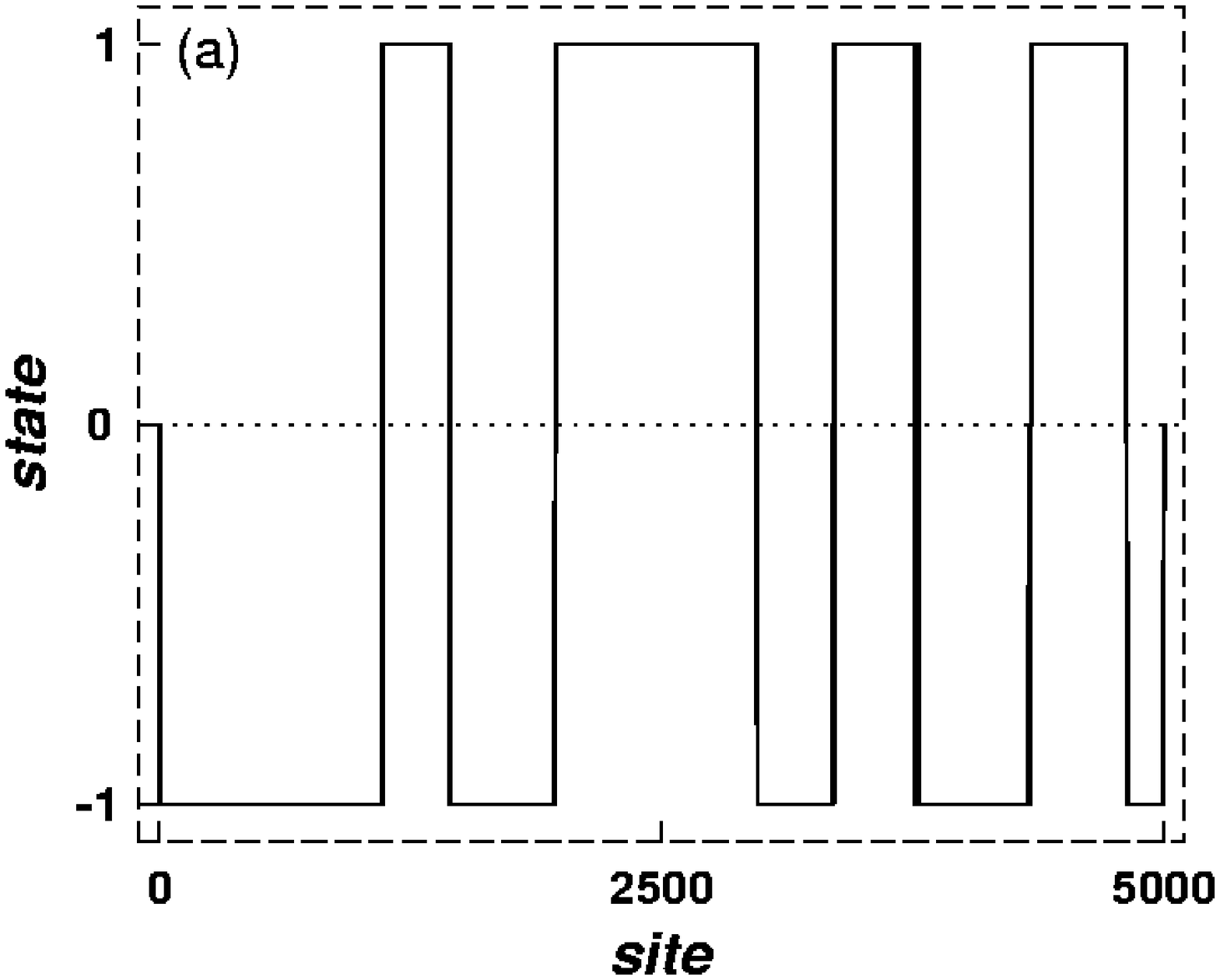}
\end{minipage}%
\begin{minipage}[c]{.5\textwidth}
\includegraphics[width=.9\textwidth]{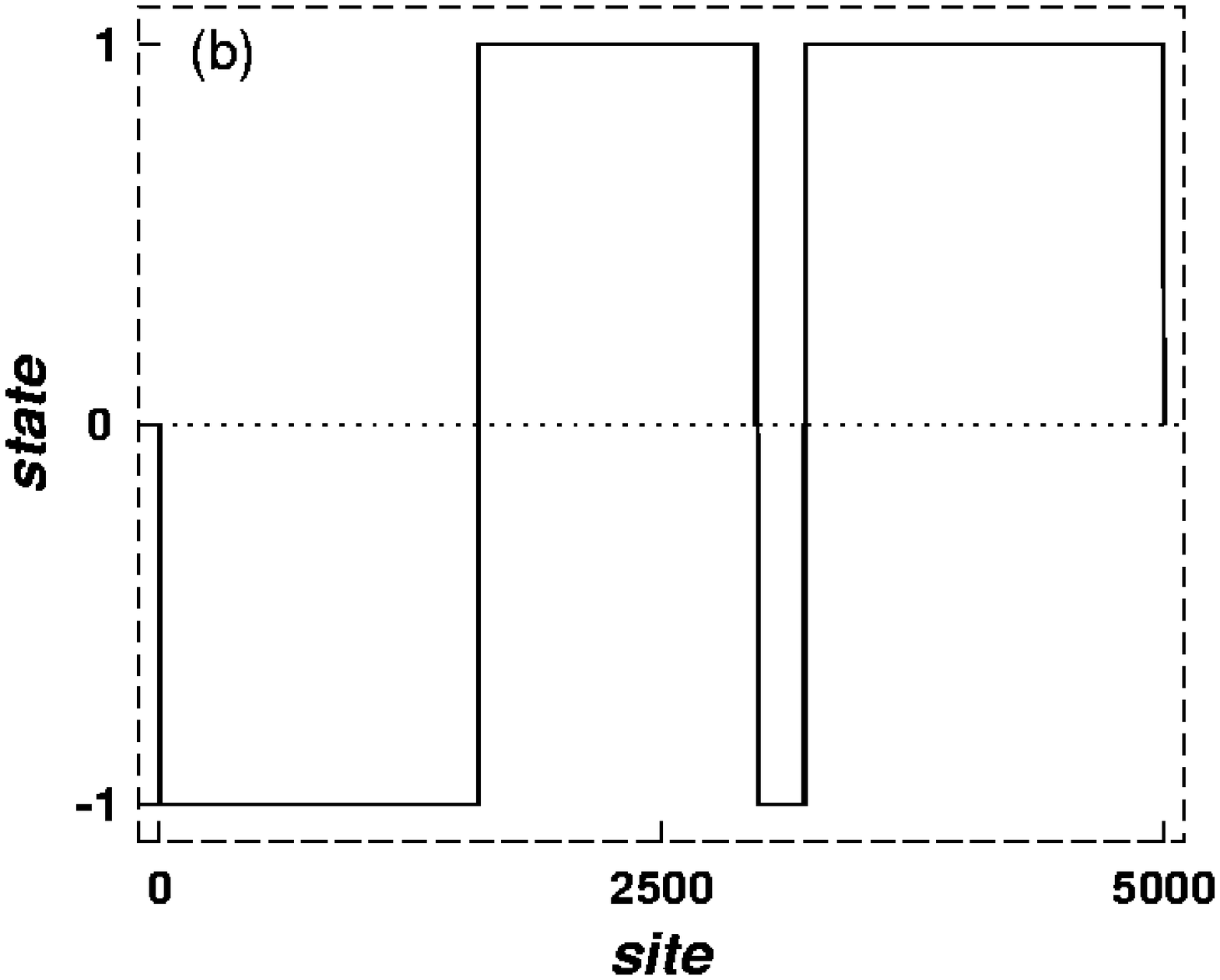}
\end{minipage}\\
\vspace{.1in}
\begin{minipage}[c]{.5\textwidth}
\includegraphics[width=.9\textwidth]{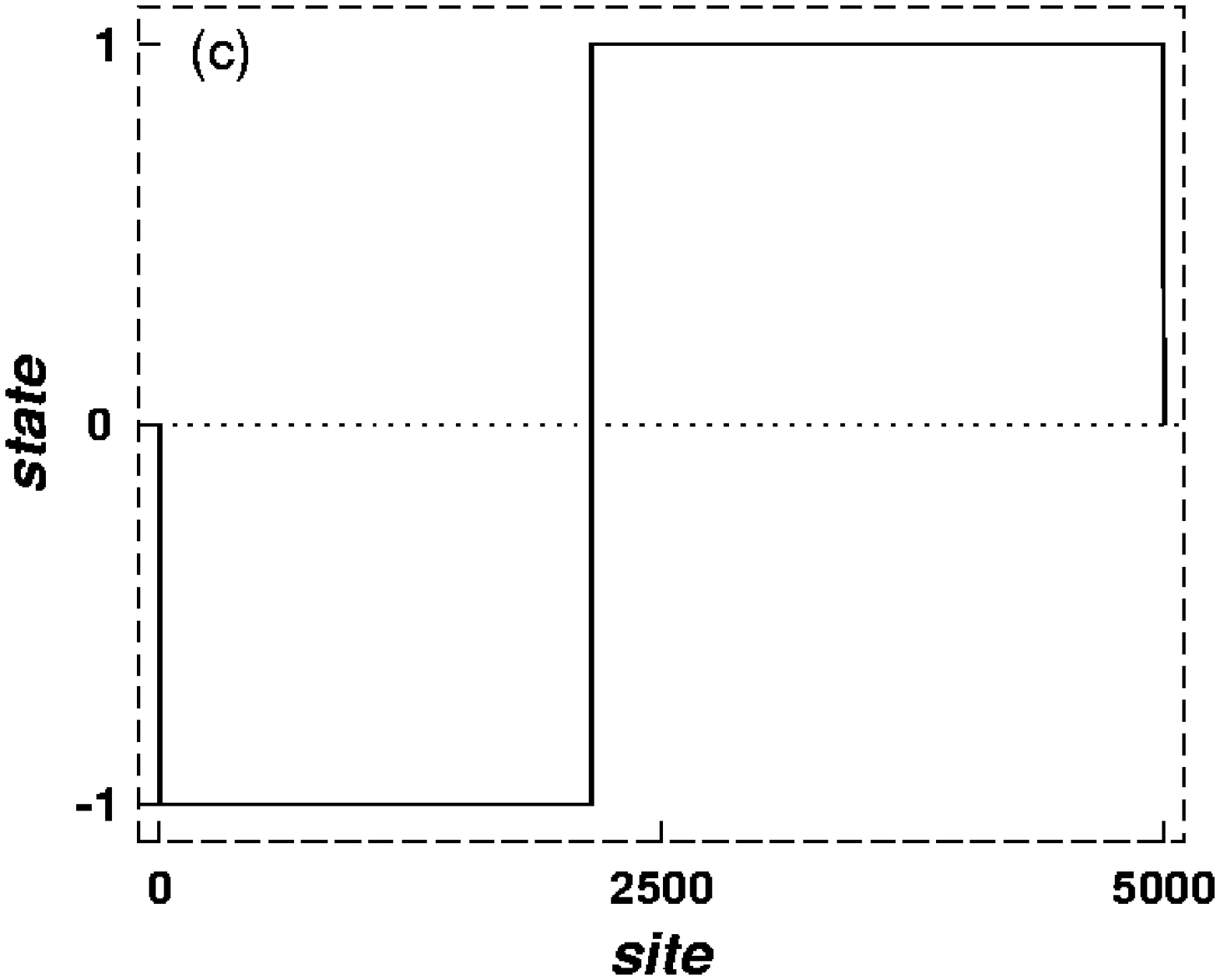}
\end{minipage}%
\begin{minipage}[c]{.5\textwidth}
\includegraphics[width=.9\textwidth]{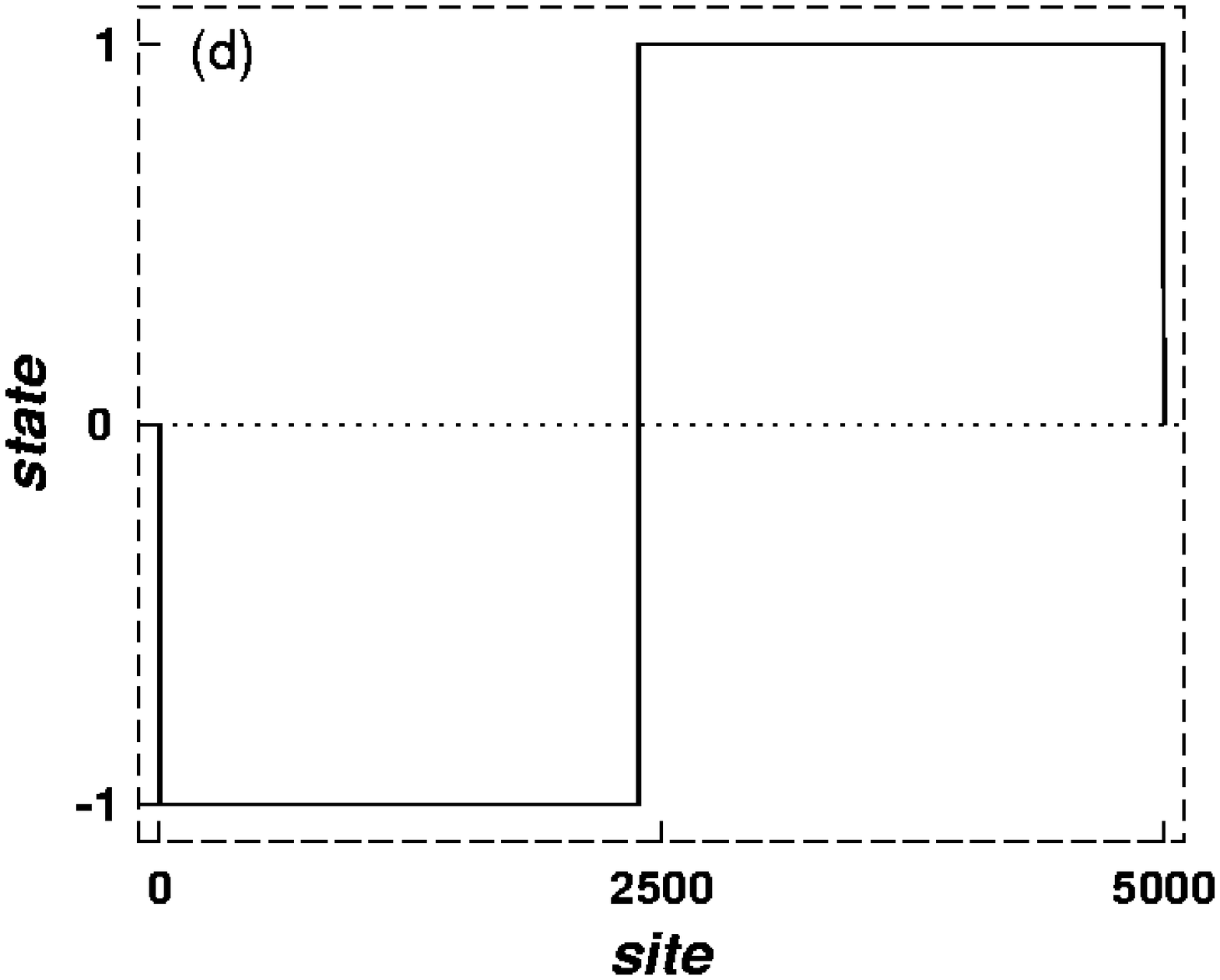}
\end{minipage}
\caption{
\label{Fig4}
{\small Occupation of a lattice of size $N = 5000$ after
(a) $2.5 \times 10^6$, (b) $5 \times 10^6$, (c) $7.5 \times 10^6$,
and (d) $10^7$ updates. Particles of type $A$ and $B$ correspond to +1
and -1, respectively.
}
}
\end{center}
\end{figure}

The result in Eq.~(\ref{na_half}) tells that in this limit the sites are on average
equally occupied by $A$ and $B$ particles. Since these findings give no insight into
the actual spatial arrangement, i.e., the correlations, we have performed a simple
Monte Carlo simulation. We start with an empty lattice of $N$ sites and update the
occupation numbers as follows. At any step, an empty (0) site is picked at random and
an $A$ (+1) or a $B$ (-1) particle is deposited with equal probabilities (corresponding
to equal activities $z_A=z_B=z$). If any of the two neighboring sites are occupied by
a particle of opposite sign, the deposited particle and the neighboring one are
removed, i.e., the occupation numbers of the sites are set to $0$. If both of the
neighboring sites are occupied by particles from the other species, one of them is
randomly selected for desorption. Desorption of individual particles is disregarded
since we are considering the limit $z \to \infty$. Fig.~\ref{Fig4} shows typical
simulation results for the evolution of the system.
The results clearly show the formation of domains consisting of mainly $A$ or $B$
particles such that only a small number, eventually just two in the limit
$N \to \infty$, of large domains emerges. Intuitively, the emergence of two domains
is due to the fact the in the limit $z \to \infty$ the number of empty sites should be
minimum since the ratio of the probability of a state with a greater number of empty
sites and the probability of the state with minimum number of empty sites goes
to zero as $z \to \infty$. Here we note that for a finite size system
increasing the number of updates will eventually lead to {\it only one} domain of
$A$ or, equally probable, $B$, spanning the whole lattice. This is due to the fact
that for $z \to \infty$ tiny deviations from $z_A = z_B$, i.e., from exactly equal
probabilities of depositing $A$ or $B$, are sufficient to drive the system into one of
the $n_A=0$ or $n_A=1$ states (see also Fig.~\ref{Fig2}). These results show that in
the limit $z_A = z_B \to \infty$, the system undergoes, in the thermodynamic limit
$N \to \infty$, segregation in a state in which half of the sites belong to a
domain of either $A$ or $B$ particles. Interestingly enough, similar behavior
has been predicted for the steady-state behavior of \textit{diffusion-limited}
$A + B \to 0$ reactions in systems with steady injection of the reactive species by
external sources with equal intensities \cite{ov,katja} (for more details see also
Ref.\cite{arg}).

\section{Conclusions.}
In this paper we have presented a model of monomer-monomer $A + B \to 0$ catalytic
reactions on a one-dimensional chain in contact with a reservoir of $A$ and $B$
particles. The model assumes continuous exchange of $A$ and $B$ species between
the chain and the vapor phase acting as a reservoir, and instantaneous reaction and
desorption of neighboring $A$ and $B$ particles. We have calculated exactly the
partition sum taking into account equilibrium fluctuations. From this we have obtained
the pressure of the adsorbed particles, the mean density, and the compressibility.
Altough in this one-dimensional model there is no phase-transition, the system
exhibits a rather non-trivial behavior. In particular, the mean density of the $A$
particles changes rapidly from very small values when the activity $z_B$ is larger
than the activity $z_A$ to a state in which the lattice is occupied predominantly
by $A$ particles when $z_A$ is larger than $z_B$. In the case when the two activities
are exactly equal and large the system undergoes segregation such that each of the
two species clusters into a large domain and occupies half of sites of the chain.

\section{Acknowledgments}
We thank Prof. P. L. Krapivsky for a critical reading of the manuscript and
helpful comments.

\end{document}